\documentclass[conference]{IEEEtran}
\IEEEoverridecommandlockouts
\usepackage{cite}
\usepackage{amsmath,amssymb,amsfonts}
\usepackage{algorithm}
\usepackage{algorithmic}
\usepackage{graphicx}
\usepackage{orcidlink}
\usepackage{textcomp}
\usepackage{amsmath}
\usepackage{booktabs}
\usepackage{xcolor}
\def\BibTeX{{\rm B\kern-.05em{\sc i\kern-.025em b}\kern-.08em
    T\kern-.1667em\lower.7ex\hbox{E}\kern-.125emX}}

\usepackage{eso-pic}
\DeclareMathOperator*{\argmin}{arg\,min}

\newcommand{\IEEEcopyrightnotice}{
\color{gray}
\textcopyright~2026 IEEE. Personal use of this material is permitted. Permission from IEEE must be obtained for all other uses, in any current or future media, including reprinting/republishing this material for advertising or promotional purposes, creating new collective works, for resale or redistribution to servers or lists, or reuse of any copyrighted component of this work in other works.
\smallskip\par
This paper has been accepted for publication in the Proceedings of the 2026 IEEE Radar Conference (RadarConf26).
}
\begin{document}

\title{A Unified Maximum-Likelihood Framework for 3D InISAR Phase Unwrapping with Outlier Rejection}

% \author{\IEEEauthorblockN{1\textsuperscript{st} Matteo Pardi\,\orcidlink{0009-0007-0615-6291}}
% \IEEEauthorblockA{\textit{University of Pisa} \\
% Pisa, Italy and\\
% \textit{RaSS National Laboratory} \\
% CNIT, Pisa, Italy}
% \and
% \IEEEauthorblockN{2\textsuperscript{nd} Francesco Mancuso\,\orcidlink{0000-0003-4174-7816}}
% \IEEEauthorblockA{\textit{RaSS National Laboratory} \\
% CNIT, Pisa, Italy}
% \and
% \IEEEauthorblockN{3\textsuperscript{rd} Elisa Giusti\,\orcidlink{0000-0001-5278-6509}}
% \IEEEauthorblockA{\textit{RaSS National Laboratory} \\
% CNIT, Pisa, Italy}
% \and
% \IEEEauthorblockN{4\textsuperscript{th} Marco Martorella\,\orcidlink{0000-0002-8985-5069}}
% \IEEEauthorblockA{\textit{University of Birmingham} \\
% Birmingham, UK and\\
% \textit{RaSS National Laboratory} \\
% CNIT, Pisa, Italy}
% }

\author{
\IEEEauthorblockN{
Matteo Pardi\,\orcidlink{0009-0007-0615-6291}\IEEEauthorrefmark{2}\IEEEauthorrefmark{1},
Francesco Mancuso\,\orcidlink{0000-0003-4174-7816}\IEEEauthorrefmark{1},
Elisa Giusti\,\orcidlink{0000-0001-5278-6509}\IEEEauthorrefmark{1},
Marco Martorella\,\orcidlink{0000-0002-8985-5069}\IEEEauthorrefmark{3}\IEEEauthorrefmark{1},
}
\vspace{3pt}
\IEEEauthorblockA{\IEEEauthorrefmark{2}Dept. of Information Engineering, University of Pisa, Pisa, Italy}
\IEEEauthorblockA{\IEEEauthorrefmark{1}Radar and Surveillance Systems (RaSS) National Laboratory, CNIT, Pisa, Italy}
\IEEEauthorblockA{\IEEEauthorrefmark{3}Dept. of Electrical and Electronic Systems Eng., University of Birmingham, Birmingham, UK}\\
\href{mailto:matteo.pardi@phd.unipi.it}{\texttt{matteo.pardi@phd.unipi.it}}
}

% --- For arXiv only ---
% Copyright overlay: first page only, in the bottom margin, no layout change.
\AddToShipoutPictureFG*{
  \AtPageLowerLeft{
    \put(\LenToUnit{(\paperwidth-\textwidth)/2 - 0.05in},\LenToUnit{0.30in}){
      \parbox[b]{\textwidth}{
        \footnotesize\IEEEcopyrightnotice
      }
    }
  }
}
% ---

\maketitle

% ===============================================================================================================
\begin{abstract}

This paper presents a novel mathematical framework for phase unwrapping in three-dimensional interferometric ISAR (3D InISAR) imaging. The approach works on a scatterer-by-scatterer basis and does not rely on any spatial continuity assumptions, making it suitable for sparse point clouds. The formulation is derived from the Mixed-Integer Least Squares (MILS) theory, an optimal maximum-likelihood framework for joint estimation of integer and real unknowns in the presence of Gaussian noise. This provides a unified way to handle generic sensor geometries, multi-baseline, multi-frequency, or hybrid setups. The method also produces a natural a posteriori quality metric for each unwrapped phase, which can be used to build a statistical test to reject outliers. The algorithm is simple to implement and has a computational cost suitable for operational systems. This paper presents the theoretical foundations of the framework and a first validation study on a standard L-shaped dual-frequency setup using Monte Carlo simulations. Results show that the proposed framework enables reliable 3D reconstruction in challenging ambiguity conditions.

\end{abstract}


% ===============================================================================================================
\section{Introduction}\label{sec:intro}

Modern radar systems are evolving toward multifunctional, multistatic, and adaptive architectures. Besides classical detection tasks, high-resolution imaging and target classification have become key capabilities. For cooperative or stationary targets, these goals are often achieved through Synthetic Aperture Radar (SAR), while for non-cooperative targets with unknown motion, the reference technique is Inverse SAR (ISAR).

Within ISAR, Three-Dimensional Interferometric ISAR (3D InISAR) has emerged as one of the most effective methods for 3D target reconstruction \cite{martorella20143d, xu2001three, tian2018review}. In its standard form, it uses three antennas in an L-shaped setup to measure interferometric phase differences and estimate the height of each scatterer relative to the Image Projection Plane (IPP). This produces a 3D point cloud, which is easier to interpret than classical 2D ISAR images and is more robust to motion uncertainties. It is also suitable for multistatic data fusion \cite{salvetti2018multiview} and automatic target recognition (ATR) \cite{jiang2023radar, meucci2025transformer, pui2023target}.

A major challenge in 3D InISAR is the \emph{phase unwrapping} problem. Because the measured phase is wrapped in the range $[-\pi,\pi)$, the height of a scatterer is known only within repeated ambiguity intervals. A common solution is to choose the radar geometry so that the whole target stays inside the unambiguous height range \cite{xu2001three, giusti2021drone}. However, this limits the system design and may not work in many real setups.

To address this issue, two active phase unwrapping strategies are usually adopted: multi-baseline and multi-frequency methods. Multi-baseline approaches increase the number of antennas \cite{ng2017long}, while multi-frequency techniques split the bandwidth into different subbands \cite{mancuso2024dual}. Both rely on diversity in ambiguity intervals to recover the correct unwrapped phase. Originating from SAR interferometry (InSAR) \cite{yu2019phase}, these techniques are suitable for 3D InISAR because they can work on a scatterer-by-scatterer basis. Other InSAR methods based on spatial continuity assumptions \cite{lin2022improved, costantini2024efficient}, or recent deep learning approaches that rely on spatial patterns \cite{yang2024unwrap}, are less suitable since 3D InISAR involves sparse point clouds.

Despite these advances, current techniques still face limitations. They often assume fixed geometries (L-shape) and uncorrelated interferometric channels, which lowers system flexibility. Moreover, they also lack a unified statistical framework and do not provide metrics for the unwrapping reliability for system design or outlier rejection.

In this work, we introduce a novel mathematical framework for phase unwrapping in 3D InISAR based on the Mixed-Integer Least Squares (MILS) theory \cite{teunissen2017springer, teunissen2006insar}. Originally developed for Global Navigation Satellite Systems (GNSS), and later applied in other fields including InSAR \cite{kampes2004ambiguity}, the MILS framework provides a statistically optimal way to jointly estimate integer ambiguities and scatterer positions according to the maximum-likelihood criterion. The method is inherently general and applicable to arbitrary sensor geometries and multi-frequency setups. In this paper, we present a first validation using a standard L-shaped configuration; broader configurations and more complex scenarios will be addressed in future research.

The main contributions of this study are as follows:
\begin{itemize}
    \item Introduce a MILS-based theoretical formulation for 3D InISAR phase unwrapping.
    \item Derive from this formulation an optimal method that is easily extendable to arbitrary sensor geometries and frequency combinations.
    \item Introduce an a posteriori metric for outlier rejection.
    \item Define performance metrics for system evaluation and optimization.
    \item Present a first validation study using a standard L-shaped configuration with dual-frequency setup.
\end{itemize}

This paper is organized as follows. Section \ref{sec:background} introduces the 3D InISAR phase unwrapping problem. Section \ref{sec:method} presents the MILS theory and the proposed method. Section \ref{sec:case_studies} describes the test scenario and the Monte Carlo simulations used to check the method’s performance. Section \ref{sec:results} shows and discusses the simulation results. Finally, Section \ref{sec:conclusion} summarizes the main points and suggests future work.

% ===============================================================================================================
\section{Background on 3D InISAR}\label{sec:background}

\begin{figure}[tb]
    \centering
    \includegraphics[width=\linewidth]{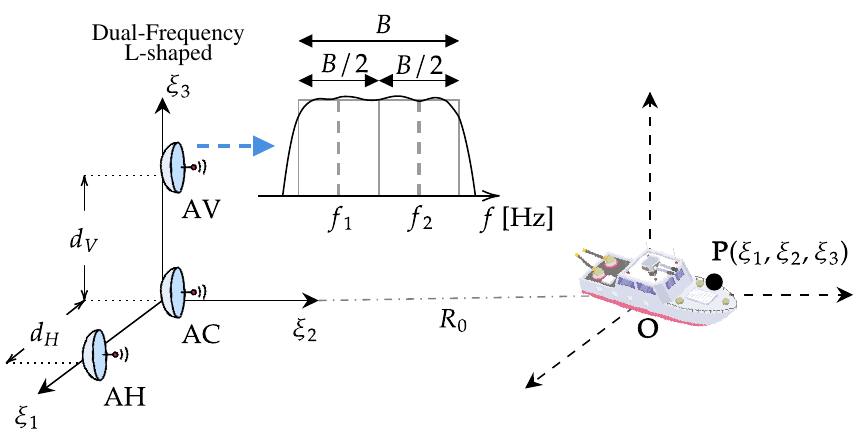}
    \caption{Schematic representation of the considered 3D InISAR system. The antennas provide vertical and horizontal baselines, and the dual-frequency approach improves phase diversity to allow unwrapped height reconstruction.}
    \label{fig:sysGeometry}
\end{figure}

Classical 3D InISAR systems employ two interferometric baselines to retrieve the full three-dimensional distribution of scatterers relative to the IPP. Figure \ref{fig:sysGeometry} illustrates a typical 3D InISAR system with three receiving antennas \{AC, AH, AV\} arranged along two orthogonal baselines and observing a target at range $R_0$. Each antenna pair, H = \{AC, AH\} and V = \{AC, AV\}, forms a baseline of length $d_H$ and $d_V$, respectively. For a non-squinted scenario, the interferometric phase measurements for each frequency channel are modeled as \cite{martorella20143d}:
\begin{equation}\label{eq:phase_model}
\begin{cases}
    \Delta \phi_H = 2 \pi k_H + \dfrac{4 \pi f}{R_0 c} d_H \xi_1, \\[6pt]
    \Delta \phi_V = 2 \pi k_V + \dfrac{4 \pi f}{R_0 c} d_V \xi_3,
\end{cases}
\end{equation}
where $f$ is the radar frequency, $c$ is the speed of light, $k_H, k_V \in \mathbb{Z}$ are the ambiguity integers, and $(\xi_1, \xi_3)$ are the scatterer coordinates in the horizontal and vertical directions (while $\xi_2$ is the down-range). The height ambiguities corresponding to each baseline are given by:
\begin{equation}\label{eq:unambiguous_height}
    H_H = \frac{R_0 c}{2 f d_H}, \qquad H_V = \frac{R_0 c}{2 f d_V}.
\end{equation}
These values represent the maximum unambiguous height for each baseline. To exceed these limits, measurements taken with different baselines or frequencies can be combined. Since $H \propto 1 / (f d)$, suitable baseline-frequency combinations provide different ambiguity intervals, and this diversity can be exploited to solve the phase ambiguity.

\subsection{General 3D InISAR Measurements Model}\label{sec:general_case}

In a general 3D InISAR system, with an arbitrary squint angle, any number of antennas, and multiple frequency subbands, the interferometric phase for the $\alpha$-th interferometric channel can be expressed as
\begin{equation}\label{eq:phase_measurement}
    \Delta \phi_\alpha = 2 \pi k_\alpha + \frac{4 \pi f_\alpha}{R_0 c} \mathbf{d}_\alpha^T \boldsymbol{\xi},
\end{equation}
where $\mathbf{d}_\alpha = [d_{\alpha 1}, d_{\alpha 3}]^T$ is the baseline vector, $f_\alpha$ is the channel's frequency, and $\boldsymbol{\xi} = [\xi_1, \xi_3]^T$. Each interferometric channel measures the projection of $\boldsymbol{\xi}$ along its corresponding baseline vector $\mathbf{d}_\alpha$, with an unambiguous height given by $H_\alpha = c R_0 / (2 f_\alpha \|\mathbf{d}_\alpha\|)$. By stacking all phase measurements into a vector $\mathbf{\Delta} \boldsymbol{\phi}$, the model (\ref{eq:phase_measurement}) can be written in matrix form:
\begin{equation}\label{eq:phase_measurement_matrix}
    \mathbf{\Delta} \boldsymbol{\phi} = 2 \pi \mathbf{k} + \frac{2 \pi}{R_0} \mathbf{D} \boldsymbol{\xi},
\end{equation}
where the elements of the matrix $\mathbf{D}$ are $D_{\alpha j} = 2 f_\alpha d_{\alpha j} / c$, with $j \in \{1, 3\}$, and $\mathbf{k}$ denotes the integer ambiguity vector.

\subsection{Phase Variance}\label{sec:phase_variance}

For each single interferometric channel, the phase variance can be modeled as \cite{esfahany2017exploitation, ferretti2007insar}:
\begin{equation}\label{eq:phase_variance}
    \sigma_\phi^2 = \frac{1 - \gamma^2}{2 \gamma^2}, \qquad \gamma = \frac{1}{1 + \frac{1}{\text{SNR}}},
\end{equation}
where $\gamma$ is known as the interferometric coherence, and depends on several factors such as the scatterer viewing angle, temporal decorrelation, and signal-to-noise ratio (SNR). In 3D InISAR, the dominant contribution is the SNR, so, for simplicity, in this analysis we model coherence only as a function of SNR. The full covariance matrix of the interferometric measurements $\mathbf{C} = \mathrm{Cov}(\mathbf{\Delta} \boldsymbol{\phi}, \mathbf{\Delta} \boldsymbol{\phi}^T)$ has the phase variance $\sigma_\phi^2$ on the diagonal, while the off-diagonal elements represent the inter-channel correlations and must be modeled according to the specific system configuration. For example, in a multi-frequency setup, interferometric channels at different frequencies can be considered uncorrelated. Instead, channels at the same frequency will be correlated if the sensor pairs share an antenna. E.g.: In Figure \ref{fig:sysGeometry}, H = \{AC, AH\} and V = \{AC, AV\} will be correlated since the antenna AC is shared.

% ===============================================================================================================
\section{Proposed Method: MILS Unwrapping}\label{sec:method}

This section describes the proposed method for solving the phase unwrapping problem in 3D InISAR.

Assuming Gaussian noise on $\mathbf{\Delta} \boldsymbol{\phi}$, the observation model (\ref{eq:phase_measurement_matrix}) can be written again by renaming the variables as
\begin{equation}\label{eq:observation_model}
    \mathbf{y} = \mathbf{A}\mathbf{a} + \mathbf{B}\mathbf{b} + \mathbf{e} \quad \mathbf{e} \sim \mathcal{N}(\mathbf{0}, \mathbf{Q}_{\mathbf{yy}}),
\end{equation}
where:
\begin{itemize}
    \item $\mathbf{y} \equiv \mathbf{\Delta} \boldsymbol{\phi} \in \mathbb{R}^{m}$ is the observation vector;
    \item $\mathbf{A} \equiv 2 \pi \mathbf{I}_{m \times m} \in \mathbb{R}^{m \times n}$ and $\mathbf{B} \equiv \frac{2 \pi}{R_0} \mathbf{D} \in \mathbb{R}^{m \times p}$ are the design matrices (in our case: $m=n$ and $p=2$);
    \item $\mathbf{a} \equiv \mathbf{k} \in \mathbb{Z}^{n}$ is the vector of integer unknowns;
    \item $\mathbf{b} \equiv \boldsymbol{\xi} \in \mathbb{R}^{p}$ is the vector of real unknowns;
    \item $\mathbf{e} \in \mathbb{R}^{m}$ is the Gaussian noise vector;
    \item $\mathbf{Q}_{\mathbf{yy}} \in \mathbb{R}^{m \times m}$ is the measurement covariance matrix.
\end{itemize}
These symbols follow the standard MILS notation \cite{teunissen2017springer} to highlight the link between the 3D InISAR model and the MILS theory. The optimal way to estimate $\mathbf{a}$ and $\mathbf{b}$ is through maximum-likelihood estimation (MLE):
\begin{equation} \label{eq:cost_function}
\begin{split}
    \mathcal{L}(\mathbf{a}, \mathbf{b} \, | \, \mathbf{y}) &:= \|\mathbf{y} - \mathbf{A}\mathbf{a} - \mathbf{B}\mathbf{b}\|_{\mathbf{Q}_{\mathbf{yy}}^{-1}}^2 \\
    \{\hat{\mathbf{a}}, \hat{\mathbf{b}}\} &= \argmin_{\mathbf{a} \in \mathbb{Z}^n,\, \mathbf{b} \in \mathbb{R}^p} \mathcal{L}(\mathbf{a}, \mathbf{b} \, | \, \mathbf{y}).
\end{split}
\end{equation}
where the notation $\|\mathbf{x}\|_\mathbf{W}^2 := \mathbf{x}^T \mathbf{W} \mathbf{x}$ is used. The quadratic form $\mathcal{L}(\mathbf{a}, \mathbf{b}\,|\,\mathbf{y})$ is proportional to the negative log-likelihood of the Gaussian distribution and acts as the cost function to minimize. The joint optimization (\ref{eq:cost_function}) is known as \emph{Mixed-Integer Least Squares} (MILS) problem, where “mixed-integer” refers to the presence of both integer and real unknowns \cite{teunissen2017springer}.

For any fixed candidate value of $\mathbf{a}$, the optimal estimate of $\mathbf{b}$ is obtained by solving $\argmin_{\mathbf{b} \in \mathbb{R}^p} \mathcal{L}(\mathbf{a}, \mathbf{b}\,|\,\mathbf{y})$:
\begin{equation}\label{eq:fixed_solution}
    \hat{\mathbf{b}}(\mathbf{a}) = (\mathbf{B}^T \mathbf{Q}_{\mathbf{yy}}^{-1} \mathbf{B})^{-1}\mathbf{B}^T \mathbf{Q}_{\mathbf{yy}}^{-1} (\mathbf{y} - \mathbf{A}\mathbf{a}).
\end{equation}
Substituting this expression (\ref{eq:fixed_solution}) into the cost function (\ref{eq:cost_function}) yields an integer least squares problem for $\mathbf{a}$:
\begin{equation} \label{eq:integer_least_squares}
\begin{split}
    \mathcal{L}(\mathbf{a}\,|\,\mathbf{y}) &:= \|\mathbf{y} - \mathbf{A}\mathbf{a} - \mathbf{B}\hat{\mathbf{b}}(\mathbf{a})\|_{\mathbf{Q}_{\mathbf{yy}}^{-1}}^2 \\
    \hat{\mathbf{a}} &= \argmin_{\mathbf{a} \in \mathbb{Z}^n} \mathcal{L}(\mathbf{a}\,|\,\mathbf{y})
\end{split}
\end{equation}
Thus, the operational algorithm consists of two steps: (i) estimate the integer vector $\hat{\mathbf{a}}$ using (\ref{eq:integer_least_squares}); (ii) estimate the real vector $\hat{\mathbf{b}}$ using (\ref{eq:fixed_solution}) with the resulting $\hat{\mathbf{a}}$.

\subsection{Search Space Constraints}\label{sec:constraints}

To obtain a practical implementation, the search space of the integer variables must be bounded. This can be done by using prior knowledge of the maximum target size, which is usually available and useful to include in the model. A constraint on the real unknowns can be imposed by requiring $\mathbf{b}$ to lie in a bounded set $\mathcal{B} \subseteq \mathbb{R}^p$. In this work, we use simple box constraints:
\begin{equation}\label{eq:box_constraints}
    |\xi_1| \le \frac{L_\mathrm{max}}{2}, \qquad |\xi_3| \le \frac{L_\mathrm{max}}{2},
\end{equation}
where $L_\mathrm{max}$ is the maximum target length, introduced here as a design parameter.

From the constraint on $\mathbf{b}$ in (\ref{eq:box_constraints}), a bounded set $\mathcal{A} \subseteq \mathbb{Z}^n$ for the integer unknowns can be derived. Using (\ref{eq:phase_measurement}) and (\ref{eq:observation_model}), each interferometric channel $\alpha$ satisfies $k_\alpha = \frac{1}{2\pi}(\Delta \phi_\alpha - \frac{4\pi f_\alpha}{R_0 c}(d_{\alpha_1}\xi_1 + d_{\alpha_3}\xi_3) - e_\alpha)$. Applying the triangle inequality gives the bound
\begin{equation}\label{eq:integer_constraints}
    |k_\alpha| \le \frac{1}{2\pi} \left(\pi + \frac{4\pi f_\alpha (d_{\alpha_1} + d_{\alpha_3}) L_\mathrm{max}}{2 R_0 c} + 5\sigma_\phi \right),
\end{equation}
where the measurement noise is assumed to be bounded by $5\sigma_\phi$.
% [FUTURE NOTE TO ADD?] ($99.9999\%$ confidence interval in a Gaussian).
% [FUTURE NOTE TO ADD?] In a typical operational scenario for 3D InISAR, the expected maximum value of $|k_\alpha|$ from (\ref{eq:integer_constraints}) is on the order of 1 (say, $\lesssim$ 10).
The set $\mathcal{A}$ is then defined as the set of all $\mathbf{a} \in \mathbb{Z}^n$ satisfying the constraint (\ref{eq:integer_constraints}) and producing a $\hat{\mathbf{b}}(\mathbf{a})$ from (\ref{eq:fixed_solution}) within $\mathcal{B}$, i.e., satisfying (\ref{eq:box_constraints}).

The operational search procedure is therefore:
\begin{enumerate}
    \item For each candidate $\mathbf{a} \in \mathbb{Z}^n$ inside the box defined by (\ref{eq:integer_constraints}), compute the cost $\mathcal{L}(\mathbf{a}\,|\,\mathbf{y})$.
    \item Select the candidate $\hat{\mathbf{a}}$ that gives the lowest cost, provided the corresponding $\hat{\mathbf{b}}$ from (\ref{eq:fixed_solution}) satisfies (\ref{eq:box_constraints}); otherwise continue to the next candidate.
    \item Finally, estimate $\hat{\mathbf{b}}$ using (\ref{eq:fixed_solution}) with the selected $\hat{\mathbf{a}}$.
\end{enumerate}
More efficient search strategies can be developed and will be considered in future work.

\subsection{A Posteriori Quality Assessment}\label{sec:posterior_quality_assessment}

Once $\hat{\mathbf{a}}$ and $\hat{\mathbf{b}}$ are estimated, the quality of the integer solution can be evaluated using the posterior probability of $\hat{\mathbf{a}}$ given the data. In Bayesian form, the \emph{posterior probability} of any candidate $\mathbf{a} \in \mathbb{Z}^n$ is given by $P(\mathbf{a} \, | \, \mathbf{y}) \propto \int_{\mathbf{b} \in \mathcal{B}} P(\mathbf{y} \, | \, \mathbf{a},\mathbf{b})\,\mathrm{d}\mathbf{b}$, where $P(\mathbf{y} \, | \,\mathbf{a},\mathbf{b}) \propto \exp\left(-\mathcal{L}(\mathbf{a},\mathbf{b}\,|\,\mathbf{y})/2\right)$. The integral is simple to evaluate under the small phase variance assumption, as in \cite{teunissen2006insar}: for $\mathbf{a} \notin \mathcal{A}$, the integrand can be approximated as zero, while for $\mathbf{a} \in \mathcal{A}$ the integration domain $\mathcal{B}$ can be approximated as $\mathbb{R}^p$, yielding a Gaussian integral easy to compute. With this approximation, the posterior probability reduces to:
\begin{equation}\label{eq:posterior_probability}
    P(\mathbf{a}\,|\,\mathbf{y}) = \frac{\exp(-\mathcal{L}(\mathbf{a}\,|\,\mathbf{y})/2)}{\sum_{\mathbf{a}' \in \mathcal{A}} \exp(-\mathcal{L}(\mathbf{a}'\,|\,\mathbf{y})/2)}.
\end{equation}
This leads to a natural a posteriori quality measure. We define the \emph{Ambiguity Posterior} (AP) of the estimated unwrapping as
\begin{equation}\label{eq:ambiguity_posterior}
    \mathrm{AP} := P(\hat{\mathbf{a}}\,|\,\mathbf{y}).
\end{equation}
The AP value lies in $[0,1]$ and is simple to compute and interpret. In Bayesian terms, it expresses how likely the estimated integer solution is correct: values near 1 indicate high confidence, while values near 0 indicate low confidence.

\subsection{Accept/Reject Criterion}\label{sec:accept_reject_criterion}

The AP value can be used to define a statistical test to accept or reject a scatterer, i.e., to detect outliers:
\begin{equation}\label{eq:accept_reject_criterion}
    \text{Accept if }\mathrm{AP} \geq \mathrm{AP}_{\mathrm{thr}},
\end{equation}
where $\mathrm{AP}_{\mathrm{thr}} \in [0,1]$ is user-defined. Among all possible quality metrics usable for the statistical test, the AP defined in (\ref{eq:posterior_probability}) and (\ref{eq:ambiguity_posterior}) can be proven to be the optimal one, using an argument similar to the Neyman–Pearson lemma \cite{verhagen2004gnss, teunissen2017springer}. The user-defined $\mathrm{AP}_{\mathrm{thr}}$ value affects the behavior of the algorithm. To evaluate its performance, we introduce the following metrics:
\begin{itemize}
    \item \emph{Acceptance Rate} (AccR): the probability of accepting a scatterer. $\mathrm{AccR} := P(\mathrm{AP} \geq \mathrm{AP}_{\mathrm{thr}})$.
    \item \emph{Conditional Failure Rate} (CoFaR): the probability that an accepted scatterer is incorrect. $\mathrm{CoFaR} := P(\hat{\mathbf{a}} \neq \mathbf{a} \,|\, \mathrm{AP} \geq \mathrm{AP}_{\mathrm{thr}})$.
\end{itemize}
These metrics depend on the chosen threshold $\mathrm{AP}_{\mathrm{thr}}$. Two strategies can be used to set it:
\begin{enumerate}
    \item \emph{Fixed-threshold}: a constant value (e.g., $\mathrm{AP}_{\mathrm{thr}} = 90\%$) is chosen. This is simple, but scatterers with different SNR values will exhibit different $\mathrm{AccR}$ and $\mathrm{CoFaR}$, since $\mathbf{Q}_{\mathbf{yy}} \propto 1/\text{SNR}$ and AP depends on $\mathbf{Q}_{\mathbf{yy}}$.
    \item \emph{Fixed-CoFaR}: a desired CoFaR (e.g., 5\%) is chosen and the corresponding $\mathrm{AP}_{\mathrm{thr}}$ for each scatterer is computed based on its SNR. This keeps the CoFaR constant across all scatterers (while $\mathrm{AccR}$ still varies with SNR). The mapping $\mathrm{AP}_{\mathrm{thr}}(\mathrm{SNR}, \mathrm{CoFaR})$ can be precomputed with Monte Carlo simulations, as shown in Section~\ref{sec:theoretical_evaluation}.
\end{enumerate}

% ===============================================================================================================
\section{Case Study and Simulation Setup}\label{sec:case_studies}

\subsection{Case Study Description}\label{sec:case_study_description}

In this section we present a case study to illustrate how the proposed approach works in practice. Although the method can handle an arbitrary number of sensors and frequency channels, here we focus on a simple 3D InISAR setup. The goal is to provide a clear demonstration of the modeling steps and perform a first evaluation of the method. More complex configurations will be analyzed in future work.

The radar system considered is a standard 3D InISAR setup with three receiving channels arranged in an L-shaped configuration, denoted as \{AC, AH, AV\} (see Figure \ref{fig:sysGeometry}). The two baselines have a length of $d_H = d_V = 2$ m. The system is non-squinted and operates at a central frequency of $10$ GHz with a total bandwidth of $B = 800$ MHz. The received bandwidth is divided into two equally spaced subbands of $400$ MHz, centered at $f_{1} = 9.8$ GHz and $f_{2} = 10.2$ GHz. This results in a dual-frequency system that provides four interferometric channels $\{\Delta\phi_{f_{1}H}$, $\Delta\phi_{f_{1}V}$, $\Delta\phi_{f_{2}H}$, $\Delta\phi_{f_{2}V}\}$, where $\Delta\phi_{f_{i}H} = \phi_{f_{i}H} - \phi_{f_{i}C}$ and $\Delta\phi_{f_{i}V} = \phi_{f_{i}V} - \phi_{f_{i}C}$ for $i \in \{1, 2\}$. Here, $\{\phi_{f_{i}C}$, $\phi_{f_{i}H}$, $\phi_{f_{i}V}\}$ are the measured phases at the three sensors \{AC, AH, AV\}, respectively, for the frequency channel $f_i$.

We assume that the phase noise is uncorrelated across different sensors and across different frequency channels, and that it has variance $\sigma_{\phi}^{2}/2$. Under this assumption, each interferometric channel has variance $\mathrm{Var}[\Delta\phi_{f_{i}H}] = \mathrm{Var}[\phi_{f_{i}H}] + \mathrm{Var}[\phi_{f_{i}C}] = \sigma_{\phi}^{2}$, and a similar expression holds for the vertical channel. Since the antenna AC appears in both interferograms H and V, the two channels share a non-zero covariance:
\begin{equation} \label{eq:covariance}
    \mathrm{Cov}(\Delta\phi_{f_{i}H},\Delta\phi_{f_{i}V}) = \mathrm{Var}[\phi_{f_{i}C}] = \frac{\sigma_{\phi}^{2}}{2}.
\end{equation}
These elements define the noise covariance matrix $\mathbf{Q}_{\mathbf{yy}}$ used by the MILS model. Hence the case of interest is:
\begin{align}\label{eq:mils_model}
    \mathbf{y} &= [\Delta\phi_{f_{1}H}, \Delta\phi_{f_{1}V}, \Delta\phi_{f_{2}H}, \Delta\phi_{f_{2}V}]^T, \\
    \mathbf{a} &= [k_{f_{1}H}, k_{f_{1}V}, k_{f_{2}H}, k_{f_{2}V}]^T, \\
    \mathbf{b} &= [\xi_1, \xi_3]^T, \\
    \mathbf{A} &= 2 \pi \mathbf{I}_{4 \times 4}, \\
    \mathbf{B} &= \frac{4\pi}{R_{0} c} \begin{bmatrix}
    f_{1} d_H & 0 \\
    0 & f_{1} d_V \\
    f_{2} d_H & 0 \\
    0 & f_{2} d_V \\
    \end{bmatrix}, \\
    \mathbf{Q}_{\mathbf{yy}} &= \sigma_{\phi}^{2} \begin{bmatrix}
        1 & 0.5 & 0 & 0 \\
        0.5 & 1 & 0 & 0 \\
        0 & 0 & 1 & 0.5 \\
        0 & 0 & 0.5 & 1 \\
    \end{bmatrix}.
\end{align}
The value of $\sigma_{\phi}^{2}$ is linked to the SNR of the considered scatterer through (\ref{eq:phase_variance}) and is assumed to be estimated from the scatterer extraction stage of the 3D InISAR processing. Note that $\sigma_{\phi}^{2}$ enters only as a scaling factor in $\mathbf{Q}_{\mathbf{yy}}$, affecting only the AP value and not the estimated values of $\hat{\mathbf{a}}$ and $\hat{\mathbf{b}}$.

The target is placed at a nominal range of $R_{0}=1.5$ km. This relatively short distance is chosen on purpose, since it produces a small unambiguous height interval ($\approx 11$ m). To constrain the search space, the maximum target size is set to $L_\mathrm{max}=200$ m, which is a reasonable upper bound for typical targets. With this setup, the constraint on the integer ambiguities from (\ref{eq:integer_constraints}) is $|k_\alpha| \leq 10$ $\forall \alpha$, thus creating a clear and challenging phase wrapping scenario.

% [FUTURE ADD-NOTE] In a typical operational scenario for 3D InISAR, the expected maximum value of $|k_\alpha|$ from (\ref{eq:integer_constraints}) is on the order of 1 (say, $\lesssim$ 10).

\subsection{Theoretical Evaluation}\label{sec:theoretical_evaluation}

One of the advantages of the proposed method is that it is fully grounded in a statistical framework. This allows us to evaluate its theoretical performance through simple Monte Carlo simulations. In this section we study how the performance metrics AccR and CoFaR vary as functions of the SNR and of the acceptance threshold $\mathrm{AP}_{\mathrm{thr}}$ for the case study. For each pair $(\text{SNR}, \mathrm{AP}_{\mathrm{thr}})$, we generate $N_{\mathrm{trials}} = 10^{5}$ samples with $\mathbf{e} \sim \mathcal{N}(\mathbf{0}, \mathbf{Q}_{\mathbf{yy}})$ and $\xi_{j} \sim \mathrm{Unif}([-L_{\max}/2,\, L_{\max}/2])$. The obtained interferometric phases $\mathbf{y}$ are then wrapped into the interval $[-\pi,\pi)$, and the corresponding integer ambiguities $\mathbf{a}$ are derived accordingly. In the results reported below, all pairs $(\text{SNR}, \mathrm{AP}_{\mathrm{thr}})$ with $\mathrm{AccR} < 10\%$ are discarded because they are not relevant for practical use and would also reduce the number of available samples for estimating CoFaR.

\begin{figure}[t]
    \centering
    \includegraphics[width=0.85\linewidth]{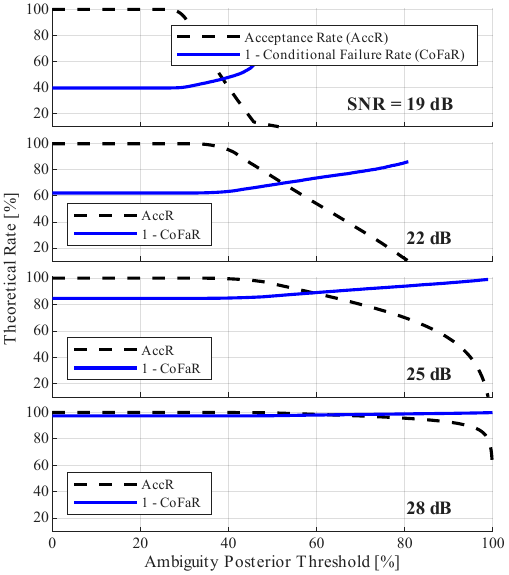}
    \vspace{-0.2cm}
    \caption{Impact of the ambiguity posterior threshold on the acceptance rate and the conditional failure rate for various SNR levels.}
    \label{fig:thr-vs-rates}
\end{figure}

Figure \ref{fig:thr-vs-rates} shows the Monte Carlo results obtained for the performance of the proposed system as a function of SNR and $\mathrm{AP}_{\mathrm{thr}}$. As expected, when $\mathrm{AP}_{\mathrm{thr}}$ is very small, all scatterers are accepted (AccR = 100\%). When $\mathrm{AP}_{\mathrm{thr}}$ increases, more scatterers are rejected, and the AccR gradually decreases until it reaches values close to zero. For a fixed $\mathrm{AP}_{\mathrm{thr}}$, the number of accepted scatterers is higher at larger SNR, because high-SNR scatterers are usually unwrapped correctly and therefore tend to produce high AP values. As also expected, increasing $\mathrm{AP}_{\mathrm{thr}}$ improves the reliability of the accepted points: the CoFaR decreases, although at the price of a smaller AccR.

\begin{figure}[t]
    \centering
    % Subfigure (a)
    \begin{minipage}[c]{0.05\linewidth}
        \centering
        \footnotesize (a)
    \end{minipage}
    \begin{minipage}[c]{0.80\linewidth}
        \includegraphics[width=\linewidth]{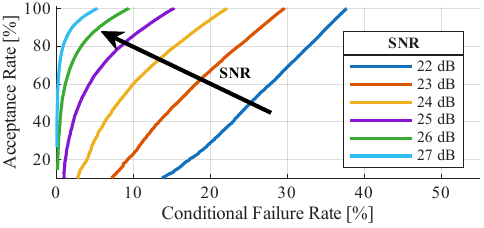}
    \end{minipage} \\
    \vspace{0.2cm}
    % Subfigure (b)
    \begin{minipage}[c]{0.05\linewidth}
        \centering
        \footnotesize (b)
    \end{minipage}
    \begin{minipage}[c]{0.80\linewidth}
        \includegraphics[width=\linewidth]{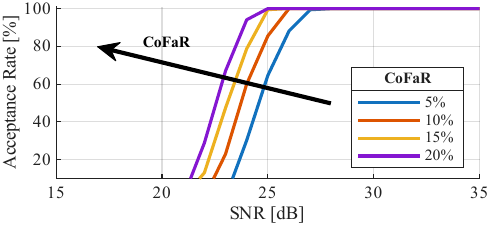}
    \end{minipage}
    % Caption
    \caption{Performance analyses: (a) ROC curves for various SNRs; (b) acceptance rate as a function of SNR for various conditional failure rates.}
    \label{fig:performance-analyses}
\end{figure}

Other useful analyses for system design are shown in Figure \ref{fig:performance-analyses}. Figure \ref{fig:performance-analyses}(a) presents the CoFaR–vs-AccR curves for different SNR values, which can be interpreted similarly to receiver operating characteristic (ROC) curves. As intuition suggests, the area under the curve increases with the SNR. Figure \ref{fig:performance-analyses}(b) shows the AccR as a function of the SNR for different fixed values of CoFaR. Here, as expected, achieving a lower CoFaR requires a higher SNR to keep the same AccR.

These analyses help the system designer understand the behavior of the algorithm and, if needed, optimize the system parameters to reach the desired performance. For example, from the presented results it is clear that the considered configuration works reliably only when the SNR is above roughly 25 dB, while performance degrades quickly for smaller values. This trend is particularly evident in Figure \ref{fig:performance-analyses}(b). It is important to note that the performance results discussed in this section refer only to the specific example considered here and should not be generalized to systems with different configurations.

% ===============================================================================================================
\section{3D InISAR Reconstruction Example}\label{sec:results}

\begin{figure}[t]
    \centering
    % Subfigure (a)
    \hspace{-1.3cm}
    \begin{minipage}[c]{0.05\linewidth}
        \centering
        \footnotesize (a)
    \end{minipage}
    \hspace{0.6cm}
    \begin{minipage}[c]{0.60\linewidth}
        \includegraphics[width=\linewidth]{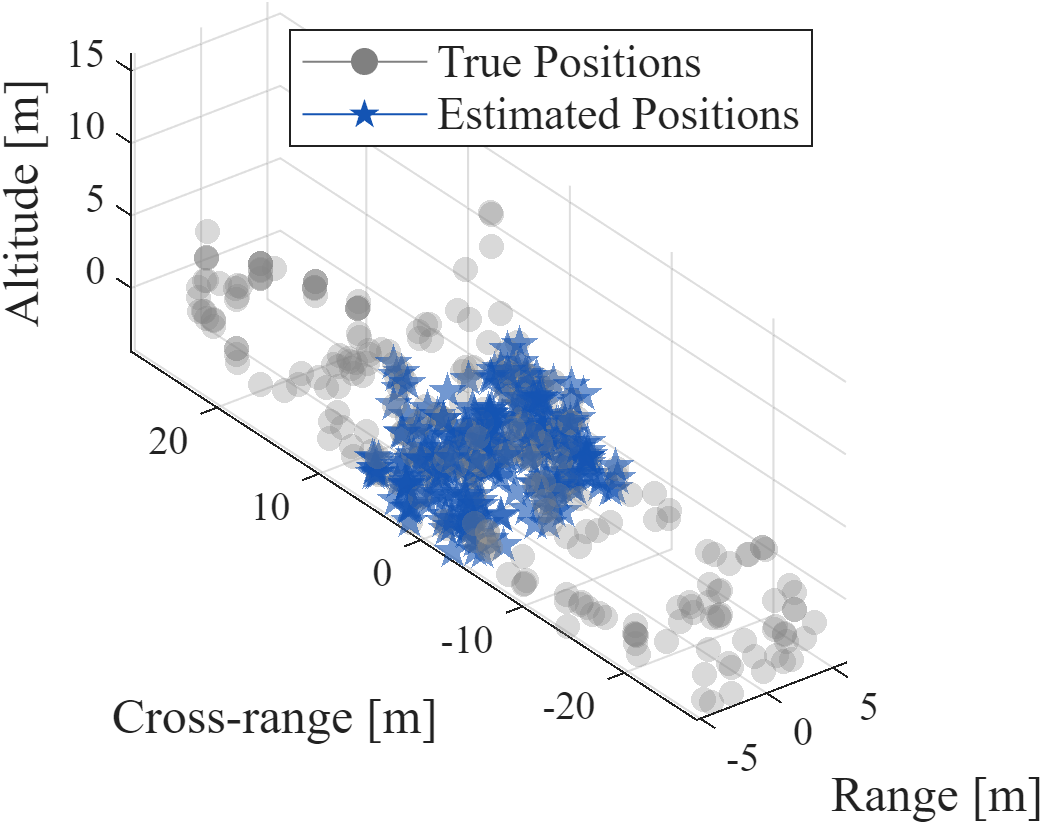}
    \end{minipage} \\
    \vspace{0.3cm}
    % Subfigure (b)
    \hspace{-0.05cm}
    \begin{minipage}[c]{0.05\linewidth}
        \centering
        \footnotesize (b)
    \end{minipage}
    \hspace{0.23cm}
    \begin{minipage}[c]{0.78\linewidth}
        \includegraphics[width=\linewidth]{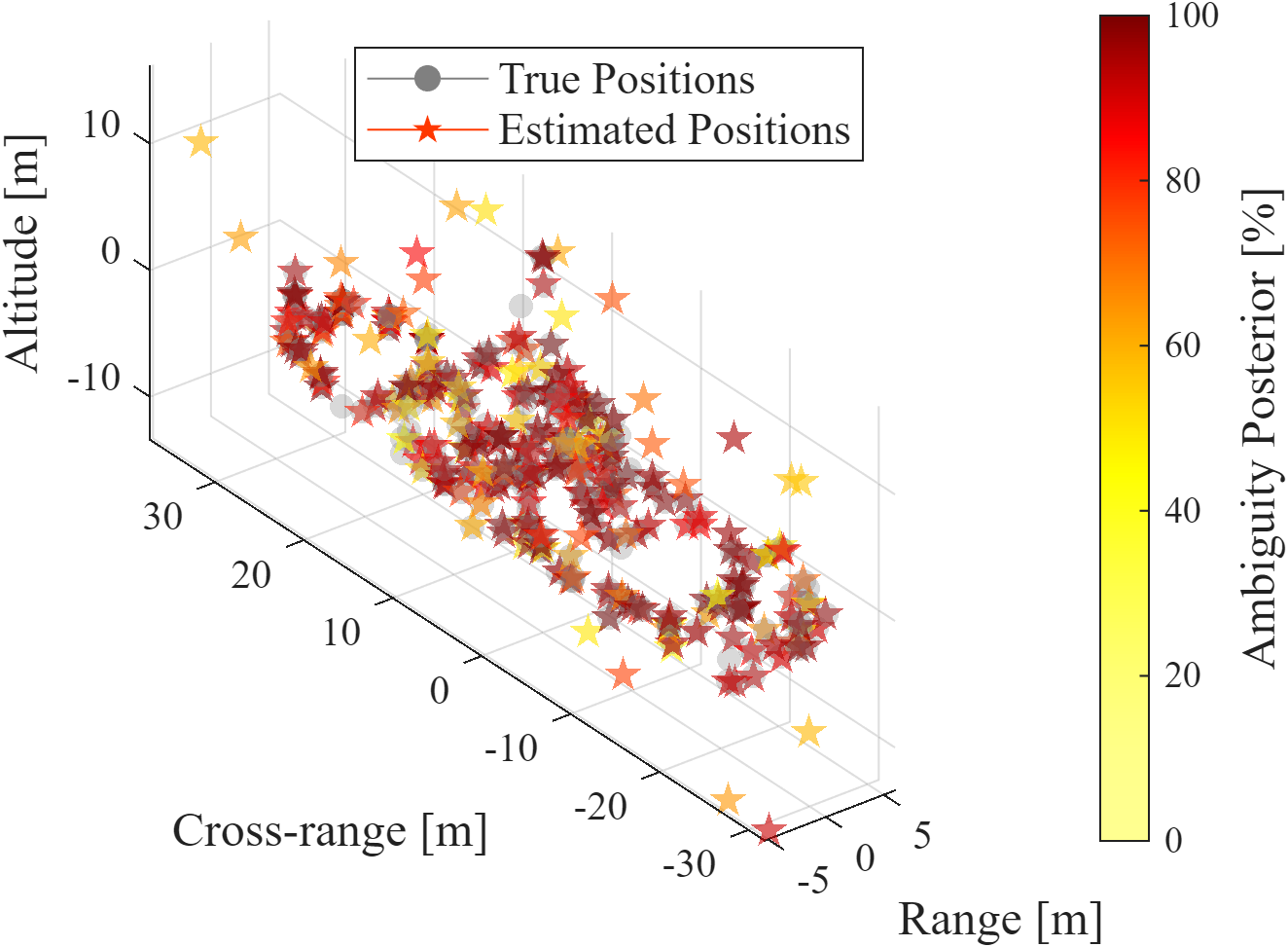}
    \end{minipage} \\
    \vspace{0.3cm}
    % Subfigure (c)
    \hspace{-0.04cm}
    \begin{minipage}[c]{0.05\linewidth}
        \centering
        \footnotesize (c)
    \end{minipage}
    \hspace{0.23cm}
    \begin{minipage}[c]{0.78\linewidth}
        \includegraphics[width=\linewidth]{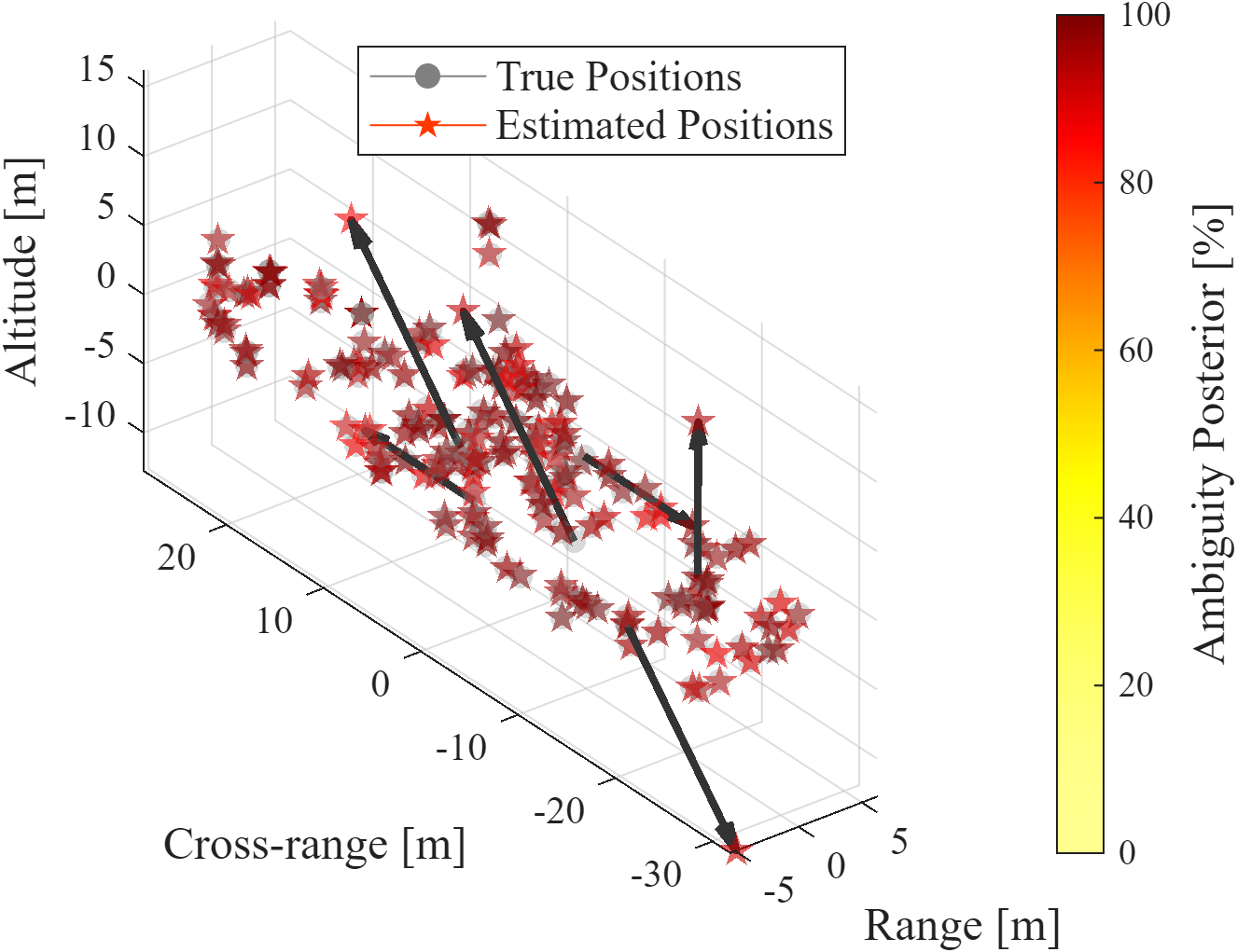}
    \end{minipage}
    % Caption
    \caption{Simulated 3D InISAR reconstruction: (a) target ground truth and reconstruction without phase unwrapping; (b) MILS reconstruction before the accept/reject stage; (c) MILS reconstruction after the accept/reject stage (CoFaR = 5\%), with incorrectly unwrapped scatterers highlighted by arrows.}
    \label{fig:3d-inisar-reconstruction}
\end{figure}

This section presents a 3D InISAR reconstruction example using a maritime target. The target is a ship measuring approximately $60 \times 10 \times 15$ m and is modeled with 312 point scatterers with a SNR of 25 dB onto the ISAR image. Figure \ref{fig:3d-inisar-reconstruction}(a) shows the true scatterer positions and the reconstruction obtained without phase unwrapping, i.e., by running the MILS algorithm while forcing all integer ambiguities to zero. (For simplicity, the range coordinate $\xi_2$ of the estimates is set to its true value, as it is not relevant for this study.) As expected, the result is severely affected by phase wrapping since the unambiguous height ($\approx 11$ m) is smaller than the extent of the target, and the reconstruction collapses and no meaningful structure can be identified.

The reconstruction obtained with MILS before the accept/reject stage is shown in Figure \ref{fig:3d-inisar-reconstruction}(b). The overall quality is significantly improved: the shape of the ship becomes visible and its approximate dimensions can be recognized. However, the result is still not satisfactory. About 11\% of the scatterers are unwrapped incorrectly, which leads to noticeable geometric distortions and inaccurate size estimates. Even so, we can see that the AP values are distributed in a way that is consistent with the expected behavior: correctly unwrapped scatterers tend to exhibit high AP values, whereas incorrectly unwrapped points typically show low AP values.

The final reconstruction, obtained after the accept/reject step, is shown in Figure \ref{fig:3d-inisar-reconstruction}(c). $\mathrm{AP}_{\mathrm{thr}}$ is selected using the fixed-CoFaR strategy: a CoFaR of 5\% is chosen, and the theoretical curves derived from Monte Carlo simulations provide the corresponding threshold $\mathrm{AP}_{\mathrm{thr}} = 84\%$ for an SNR = 25 dB. After applying this threshold, the percentage of incorrectly unwrapped scatterers decreases to about 3\%, producing a reliable 3D reconstruction of the target. The cost of this improvement is that 35\% of the scatterers are discarded, but this is acceptable to obtain a stable and accurate point cloud.

A small number of incorrectly unwrapped scatterers still remains. These residual outliers may affect a target classifier that is not robust to sparse erroneous points. In Figure \ref{fig:3d-inisar-reconstruction}(c), these scatterers are highlighted with arrows starting at the true scatterer position and ending at the estimated (incorrect) one. A possible post-processing strategy to further reduce their impact is to apply an additional outlier detection step based on local point density, since outliers are usually located in isolated or low-density regions of the reconstructed point cloud.

Table \ref{tab:rmse_dual_final_wide} summarizes the reconstruction and validation results for the target. It shows the percentage of accepted scatterers, the percentage of correctly unwrapped scatterers (among those accepted), and the Root Mean Square Error (RMSE) of the reconstructed 3D position. It is important to note that the RMSE computed only on the correctly unwrapped scatterers is much lower than the RMSE computed on all scatterers. For all three methods in the table, the RMSE computed only on the correctly unwrapped scatterers is 0.10 m.

% ===============================================================================================================
\section{Conclusion and Future Work}\label{sec:conclusion}

In this work, we introduced a novel mathematical framework for 3D InISAR phase unwrapping based on Mixed-Integer Least Squares theory, and we carried out a first evaluation through a simulated case study. The results show that the proposed approach can achieve reliable 3D reconstruction even in challenging ambiguity conditions. A key advantage of the method is the availability of an a posteriori quality metric, the ambiguity posterior, which gives the Bayesian probability that the estimated unwrapping is correct. In practice, it proved effective in improving the final reconstruction by enabling a statistical test to accept or reject a scatterer, while offering full control over the failure rate of this stage.

Several future developments are possible. For example:
\begin{itemize}
    \item Designing smarter and more efficient strategies to define the search space of the integer unknowns. This would make the approach scalable to systems with many interferometric channels, i.e., having a large number of sensors and frequency channels.
    \item Studying the AccR, CoFaR, and RMSE metrics for different 3D InISAR systems with various geometries, baselines, and frequency subbands, to find robust and optimal solutions for different operational scenarios.
    \item Assessing 3D InISAR reconstruction quality using more advanced and realistic simulations, with targets composed of scatterers having different SNRs following statistical distributions, and ultimately simulating a complete end-to-end 3D InISAR processing chain.
\end{itemize}

\begin{table}[t]
    \centering
    \caption{Reconstruction metrics for the simulated target (312 total scatterers): accepted scatterers in the accept/reject (A/R) stage, correctly unwrapped scatterers, and 3D position RMSE.}
    \begin{tabular}{lccc}
    \toprule
    Method & Accepted [\%] & Correct [\%] & RMSE [m] \\
    \midrule
    No Unwrapping & 100 & 40 & 14.3 \\
    MILS Before A/R & 100 & 89 & 4.2 \\
    \textbf{MILS After A/R} & \textbf{65} & \textbf{97} & \textbf{2.4} \\
    \bottomrule
    \end{tabular}
    %\vspace{0.055cm}
    \label{tab:rmse_dual_final_wide}
\end{table}

% ===============================================================================================================
\bibliographystyle{IEEEtran}
\bibliography{references}

\end{document}